\documentclass[pdflatex,sn-nature,oneside]{sn-jnl}

\usepackage{graphicx}
\usepackage{multirow}
\usepackage{amsmath,amssymb,amsfonts}
\usepackage{amsthm}
\usepackage{mathrsfs}
\usepackage[title]{appendix}
\usepackage{xcolor}
\usepackage{textcomp}
\usepackage{manyfoot}
\usepackage{booktabs}
\usepackage[version=4]{mhchem}
\usepackage{tabularx}
\usepackage{algorithm}
\usepackage{algorithmicx}
\usepackage{algpseudocode}
\usepackage{listings}

\hypersetup{colorlinks=true,linkcolor=blue,citecolor=blue,urlcolor=blue,hypertexnames=false}

\usepackage{placeins}
\makeatletter
\let\sn@old@section\section
\renewcommand{\section}{\FloatBarrier\sn@old@section}
\let\sn@old@subsection\subsection
\renewcommand{\subsection}{\FloatBarrier\sn@old@subsection}
\makeatother

\newcommand{\todo}[1]{\textcolor{red}{[TODO]}}

\theoremstyle{thmstyleone}

\theoremstyle{thmstyletwo}

\theoremstyle{thmstylethree}

\begin{document}

\title[LLPR active learning for MLFF training and fine-tuning]{Full-data accuracy with fewer labels for training and fine-tuning machine-learning force fields}

\author*[1,3]{\fnm{Sheng} \sur{Bi}}\email{sheng.bi@xmu.edu.cn}
\author[2,3]{\fnm{Yi-Ze} \sur{Wang}}
\author*[2,3]{\fnm{Jun} \sur{Cheng}}\email{chengjun@xmu.edu.cn}

\affil*[1]{\orgdiv{College of Materials}, \orgname{Xiamen University}, \orgaddress{\city{Xiamen}, \country{China}}}
\affil[2]{\orgdiv{State Key Laboratory of Physical Chemistry of Solid Surfaces, iChEM, College of Chemistry and Chemical Engineering}, \orgname{Xiamen University}, \orgaddress{\city{Xiamen}, \country{China}}}
\affil[3]{\orgdiv{Laboratory of AI for Electrochemistry (AI4EC)}, \orgname{IKKEM}, \orgaddress{\city{Xiamen}, \country{China}}}

\abstract{Machine-learning force fields (MLFFs) are reliable only near their training distribution, making efficient construction of diverse training sets a major bottleneck for both train-from-scratch and foundation fine-tuning workflows. Active learning can reduce this cost, but standard model-committee uncertainty is impractical for foundation MLFFs because each committee member requires a separate fine-tuning run. We present an active-learning workflow based on last-layer-projection regression (LLPR), a forward-pass-cheap per-configuration uncertainty estimator. Across molecular, condensed-phase, and electrolyte systems, LLPR identifies compact, high-value training sets that recover full-data accuracy using only a small fraction of electronic-structure labels. In foundation-model fine-tuning, LLPR-selected configurations approach the practical fine-tuning ceiling with substantially fewer labels than random selection. In iterative electrolyte fine-tuning, LLPR detects unphysical local coordination before new labelling, calibrates uncertainty in force-error units, and supports retrospective convergence monitoring. The resulting models reproduce reference density and Li--O coordination structure, providing a scalable uncertainty-quantification strategy across MLFF training regimes.}


\maketitle

\section*{Introduction}

Machine-learning force fields (MLFFs) are widely used for atomistic simulation at near density-functional-theory (DFT) accuracy and classical-molecular-dynamics throughput\cite{ref5,ref6,ref7}. Their accuracy and reliability depend strongly on the training data: an MLFF is accurate on configurations close to its training distribution, but its predictions are usually weakly constrained on configurations outside it\cite{ref38,ref39,ref41}. Two training paradigms now coexist in the field. The traditional \emph{train-from-scratch} (TFS) approach trains an MLFF on a system-specific dataset, typically a few hundred to a few thousand DFT-labelled configurations of the target chemistry\cite{ref41,ref42}. The more recent \emph{fine-tuning} (FT) approach\cite{ref8,ref9,ref10,ref11,ref43,ref44} starts from a foundation MLFF pretrained on a large and chemically diverse structure collection and adapts it to a target system using a much smaller fine-tuning dataset. The data-quality and out-of-distribution issues that affect MLFFs in general persist in the FT setting and can be more severe: a foundation has seen such a wide range of chemistries that it tends to return confident-looking predictions even far from the target system, so when the model is in fact extrapolating the failure is easy to miss\cite{ref40,ref41,ref43}.

A trained MLFF needs to do two things well, regardless of training paradigm. The first is the conventional training criterion: the energy and force errors on a held-out test set are sufficiently small, indicating that the network parameters are well fit to the labelled data\cite{ref35}. The second, and arguably more important, is that once deployed in molecular dynamics the model is both \emph{stable}, in the sense that the trajectory does not crash or develop numerical instability, and \emph{reliable}, in the sense that the configurations sampled and the observables predicted are consistent with what an MD run at the labelling level of theory would produce\cite{ref36,ref37,ref40,ref41}. The training and test sets are pre-constructed and represent only a snapshot of the relevant chemical space; once deployed in MD the model is free to explore configurations that may not appear in either set, and where it must extrapolate, the predicted potential energy surface is not constrained to be accurate\cite{ref38,ref39,ref41}. Thus, even an MLFF with low test-set MSE can produce unstable or unphysical MD when thermal fluctuations sample configurations outside the training distribution\cite{ref36,ref40}. The generalisation gap between test-set error and downstream-task reliability is particularly acute in MLFFs, because the configurational space of a chemical system is vastly larger than any practical labelled dataset and MD actively probes that space at every timestep.

What ultimately matters is the dataset itself. At a fixed budget of $N$ labelled configurations, a dataset that spreads its frames across the relevant region of configuration space generalises better in MD than one that concentrates the same number of frames in a narrow region. The central question for MLFF data generation is therefore not how much data to generate but which data\cite{ref28,ref41,ref42}. A principal approach for assembling such a dataset is \emph{active learning} (AL)\cite{ref12,ref16,ref17,ref18,ref23,ref42}, an iterative procedure in which an uncertainty quantification (UQ) signal computed by the model itself drives configuration selection. A standard active-learning workflow, implemented in tools such as DPGEN\cite{ref29} and AI$^2$-kit\cite{ref30} as well as in many in-house pipelines, alternates exploratory MD with the current model, uncertainty-ranked frame selection, DFT labelling, and retraining, until the model is judged converged.

The most common UQ signal in MLFF active learning is the \emph{model committee}\cite{ref12,ref16,ref17,ref18}, in which several MLFFs are trained from independent random initialisations on the same data and the spread of their predictions on a query configuration is taken as the uncertainty estimate\cite{ref17,ref19}. The model committee is conceptually simple and transfers across MLFF architectures, but it has a clear cost: every additional committee member is a full extra training run. The overhead grows with model size and with training-dataset size, and becomes severe at foundation MLFF scale: training, for example, four foundation copies merely to obtain a UQ signal on a target system is rarely a practical operating point. Several alternative UQ signals have been proposed in recent work on foundation MLFFs, including dedicated \emph{confidence-head} outputs trained alongside the main potential\cite{ref31}, \emph{heterogeneous ensembles} that aggregate predictions across diverse pretrained foundations\cite{ref32}, \emph{quantile-regression} heads that estimate prediction intervals directly\cite{ref33}, and environment-dependent post-hoc calibration that improves uncertainty--error correlation and high-error-configuration detection\cite{ref45}. Among these, the \emph{last-layer-projection regression} (LLPR) framework\cite{ref20} is particularly attractive: it uses the model's own last-layer features together with the covariance over the training set to estimate per-configuration uncertainty in a single forward pass. LLPR was first applied to MLFFs in the train-from-scratch setting on QM9 and bulk water\cite{ref20}; recent flexible-calibration work has also used LLPR as a baseline UQ signal for MACE-MP-0 and demonstrated high-error detection and active fine-tuning\cite{ref45}. What remains unestablished is a unified LLPR workflow that spans both train-from-scratch and foundation fine-tuning while combining pre-labelling pathology detection, iterative data selection, and retrospective convergence assessment in calibrated error units.

In this work we implement and integrate LLPR-based UQ signals into active-learning and fine-tuning workflows, and demonstrate that, with appropriate design choices, an LLPR-driven picker approaches full-data reference accuracy with substantially fewer target-system labels than random selection in both train-from-scratch and fine-tuning settings, while remaining competitive with the model committee in the from-scratch comparison. LLPR uncertainty also identifies unphysical configurations sampled during otherwise stable foundation-model MD, a failure mode that conventional stability diagnostics cannot detect. We demonstrate the workflow across three settings: train-from-scratch active learning on an open water/ice dataset\cite{ref26,ref27}, one-shot fine-tuning of MACE-MP-0 on the same water/ice target, and iterative active fine-tuning of an OMAT-small foundation\cite{ref9} on a 21 molal LiTFSI water-in-salt electrolyte for which the initial foundation MD is stable but structurally hallucinated. In the electrolyte case, calibration against pseudo-oracle labels also enables retrospective convergence assessment.

\section*{Results}

\subsection*{LLPR-based uncertainty quantification in MLFFs}
To validate that LLPR gives quantitatively calibrated uncertainty for an MLFF and to assess its effectiveness as a picker for active learning, we evaluate it using four DeePMD models and a 50-configuration reference set of 64-water cells. After excluding every geometry in that reference set and removing exact duplicate evaluation geometries, the evaluation pool comprises 1646 configurations: 585 in-distribution 64-water configurations, 1022 out-of-distribution 63-water configurations, and 39 out-of-distribution 128-water configurations, with the OOD subsets deliberately chosen as cell-size shifts of the training-set chemistry. The water configurations are drawn from the dataset of Fu \emph{et al.}\cite{ref48}, and full details of the model and the data partition are in Methods. From the model's last-layer features and the covariance over the 50-configuration reference set we compute two LLPR uncertainty signals: a per-frame \emph{energy LLPR} $\sigma^2_E$ that scores the uncertainty of the model's predicted total energy of a frame, and a \emph{top-5 force LLPR} $\sigma^2_F$ that scores its local force uncertainty as the per-frame mean of the five largest per-atom uncertainties (for details see Methods); averaging the five largest atomic uncertainties reduces sensitivity to fluctuations in the most-uncertain atom. $\sigma^2_E$ is therefore a global per-frame signal, $\sigma^2_F$ a localised per-atom signal aggregated to a frame. For comparison we also compute committee disagreement: the spread of energy and force predictions across four MLFFs trained from independent random seeds, the standard MLFF active-learning UQ baseline.

\begin{figure}[!htbp]
\centering
\includegraphics[width=\textwidth]{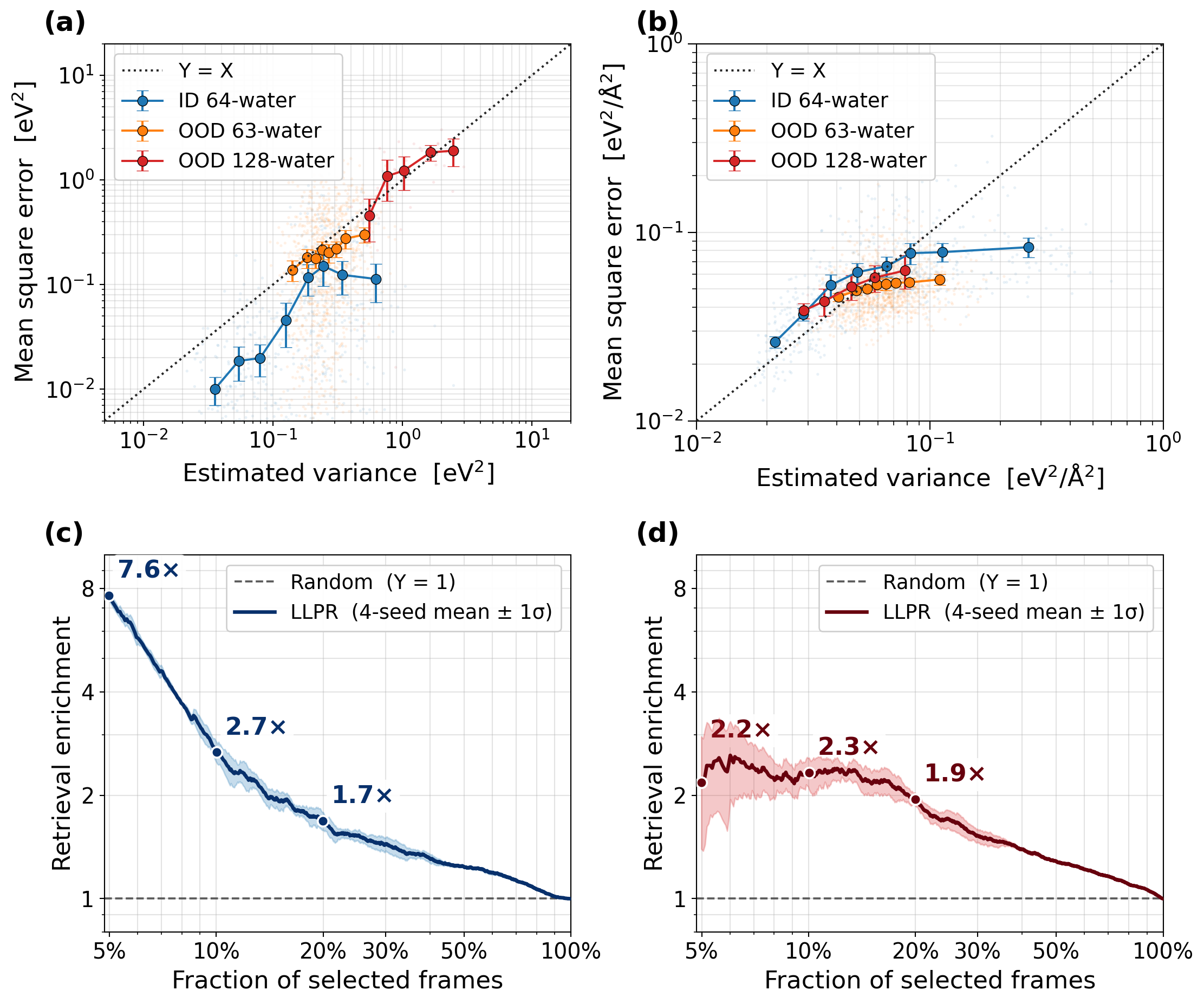}
\caption{\textbf{LLPR-based UQ validation on bulk water.} \textbf{(a, b)} Bin-mean predicted $\sigma^2$ of energy LLPR (a) and top-5 force LLPR (b) plotted against bin-mean observed squared error, shown as coloured bin-mean curves with $\pm$95\% CI whiskers. The faint scatter behind the curves is the per-group raw $(\sigma^2, e^2)$ point cloud. \textbf{(c, d)} Retrieval enrichment over random sampling for energy (c) and top-5 force (d). Random baseline = y = 1 (dashed grey). All panels use the same 1646 unique evaluation geometries, excluding the complete 50-configuration reference set. Calibration uses the seed-42 LLPR signal and committee-mean prediction error. Retrieval LLPR is shown as the mean $\pm$ one sample standard deviation across four training seeds; enrichment values at the integer budgets nearest K/N = 5\%, 10\%, 20\% are labelled on each LLPR curve.}
\label{fig:2}
\end{figure}

Both energy LLPR ($\sigma^2_E$) and force LLPR ($\sigma^2_F$) correlate strongly with the actual prediction error and provide quantitative uncertainty estimates. Figure~\ref{fig:2}a and Figure~\ref{fig:2}b plot the bin-mean predicted $\sigma^2$ against the bin-mean observed squared error for the energy and force signals respectively, in equal-count bins of the 1646-frame pool with the $y = x$ diagonal as the reference for exact agreement. A frame with high LLPR $\sigma^2$ shows a correspondingly larger actual error on average. The LLPR signals can therefore be used as quantitative uncertainty estimates and not merely as a ranking statistic.

Resolving the bin-mean curves by group reveals a clear ID/OOD contrast in energy LLPR but not in force LLPR. In Figure~\ref{fig:2}a, the in-distribution 64-water frames give the lowest $\sigma^2_E$, the 63-water OOD frames give moderately higher values, and the 128-water OOD frames give the highest $\sigma^2_E$. This ordering is expected: the model was trained on 64-water cells, so 63- and 128-water cells are both extrapolations, but 128-water is the more extreme one in terms of total system size. In Figure~\ref{fig:2}b, by contrast, the per-atom force-uncertainty distributions of the three groups overlap rather than separate. Force LLPR is a strictly local signal, so it does not respond to cell-size shifts as long as the local water environment around each atom looks the same across 64-, 63-, and 128-water cells at roughly the same density. Energy LLPR aggregates over all atoms of a frame and therefore does respond to total-system-size differences that leave the local environments intact.

To further test whether these signals translate into effective pickers for active learning, we measure their retrieval performance, defined as how often a top-K pick by LLPR-$\sigma^2$ coincides with the top-K worst-error frames (i.e., frames with the largest true prediction error), and compare it against random sampling as the no-signal baseline. Figure~\ref{fig:2}c and Figure~\ref{fig:2}d plot the \textit{enrichment factor} for energy LLPR and force LLPR respectively, defined as the fraction of worst-K frames recovered by LLPR divided by the fraction a random picker recovers at the same budget ratio K/N (K = number of picks, N = size of the pool). For energy-LLPR retrieval (Figure~\ref{fig:2}c), the four-seed mean enrichment over random is 7.6 at K/N $\approx$ 5\%, 2.7 at 10\%, and 1.7 at 20\%. For force-LLPR retrieval (Figure~\ref{fig:2}d), the corresponding enrichments are 2.2, 2.3, and 1.9. All methods use the same evaluation pool, with duplicate configurations and all reference-set configurations excluded; each nominal budget fraction is rounded to the nearest integer number of configurations. The curves approach unity when the whole pool is selected. Shaded bands show one sample standard deviation across the four training seeds and quantify the seed dependence of LLPR retrieval on this fixed pool.

\subsection*{Validating LLPR in an end-to-end active-learning workflow}

The validation above showed that LLPR's energy and force uncertainties both agree quantitatively with the true prediction error and are effective as pickers. We now test whether this picker performance transfers to an end-to-end active-learning workflow that produces a train-from-scratch MLFF, and how the resulting models compare against a from-scratch full-data ceiling.

To make the comparison fair we designed a controlled active-learning experiment on the water/ice dataset from Morawietz \textit{et al.}\cite{ref26,ref27}, which contains 7241 RPBE-D3 configurations of liquid water and ice polymorphs ranging from low-density liquid to compressed ice. Training an MLFF on this dataset is substantially more challenging than training one for liquid water alone, because the model must simultaneously capture a disordered liquid phase and several distinct crystalline phases with remarkably different densities, coordination environments, and hydrogen-bond topologies, all within a single set of network parameters. A controlled design holds everything except the UQ method constant: each strategy draws from the same pre-labelled candidate pool in each round, the cumulative picks are used to retrain the model under a fixed recipe, and every model is evaluated on the same held-out test set. Two reference baselines are essential to this comparison: random sampling, against which any improvement tells us that the UQ signal is what helps, not merely that the dataset is growing larger; and a from-scratch full-data ceiling, defined as a same-architecture MLFF trained on the full 3955-frame reference training set under the same recipe.

The bootstrap model is a DeePMD potential trained on the liquid-water subsystem of the Morawietz dataset. A multi-phase test set of 990 configurations is held out for evaluation: 500 additional liquid frames from the same subsystem plus 490 ice frames spanning the four ice-phase groups (ice\_low, ice\_mid, ice\_XV and ice\_VIII, as defined in Methods). The active-learning candidate pool from which each strategy picks consists of 1140 ice frames, namely the remaining 70\% of those four ice-phase groups not held out for testing; every ice configuration in this pool is out-of-distribution for the bootstrap model by construction.

\begin{figure}[!htbp]
\centering
\includegraphics[width=\textwidth]{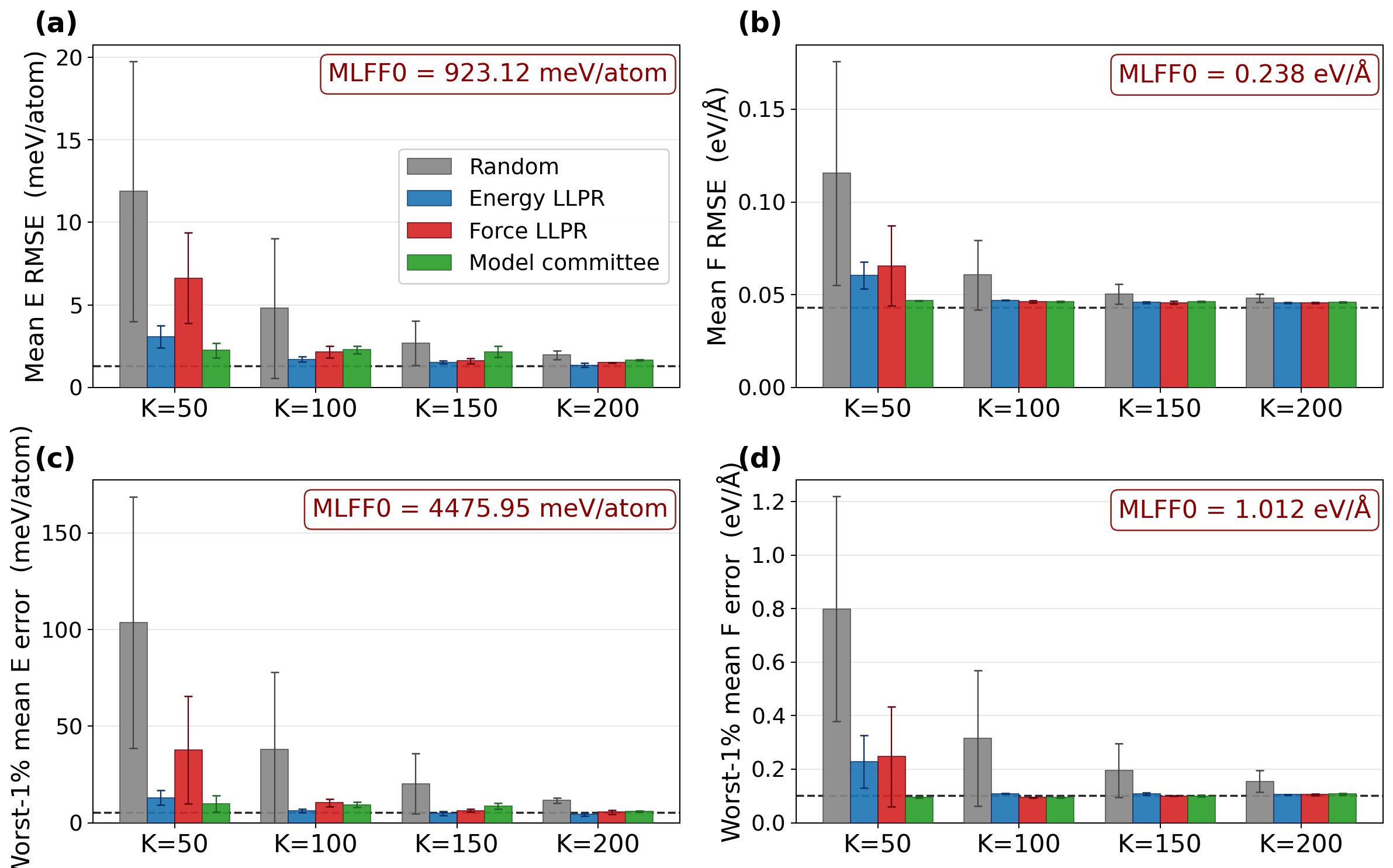}
\caption{\textbf{LLPR-based picking approaches full-data accuracy on water/ice with fewer labels.} Four picker strategies are compared at $K\in\{50,100,150,200\}$ cumulative ice labels. \textbf{(a, b)} Frame-weighted aggregate energy and force RMSE on the 990-frame test set, using three retraining seeds per cell. \textbf{(c, d)} Mean of the largest nine per-frame energy or force errors ($\lfloor990/100\rfloor$, approximately 1\%). These panels use three seeds for every cell. Error bars are population standard deviations across the three seeds (divisor $n$). The dashed lines and MLFF0 annotations give the full-data ceiling and bootstrap.}
\label{fig:3}
\end{figure}

The bootstrap model is reasonably accurate on the liquid partition of the test set, with a mean energy RMSE of about 2.83 meV/atom and a mean force RMSE of about 0.1239 eV/\AA{} (see SI). Its worst-1\% mean energy and force errors, however, are approximately 4476 meV/atom and 1.01 eV/\AA{} respectively (Figures~\ref{fig:3}c,~\ref{fig:3}d), showing that the bootstrap is severely inaccurate on the hardest test frames. The from-scratch full-data ceiling is a model trained under the same recipe on all the small-cell training data available; this ceiling reaches a worst-1\% mean energy error of 5.4 meV/atom and a worst-1\% mean force error of 0.102 eV/\AA{}, roughly three orders of magnitude better on energy and one order of magnitude better on force than the bootstrap. Per-subsystem errors of the bootstrap and the full-data ceiling are tabulated in Table S1.

We then test the two LLPR uncertainty signals as picking strategies in this workflow, together with random sampling and the model committee as reference baselines. Energy LLPR and force LLPR are the two signals defined in the previous subsection; the model committee here is the per-frame mean of the five largest per-atom force-spread values across an ensemble of independently trained MLFFs (four bootstrap members initially, then three members during iterative acquisition). For each strategy and each label count K, we train a single model on a 200-frame liquid anchor (a random subset of the bootstrap's liquid-water training set) together with the K frames picked by that strategy from the candidate pool. Every retrained model is warm-started from the bootstrap weights and trained under the same annealed schedule, so the comparison between strategies isolates which K ice picks were added. We retrain at four label counts, K = 50, 100, 150 and 200, with three independent retraining seeds at each K. The from-scratch workflow acquires five new frames per round by score ranking while excluding previous picks. The cumulative selections are then used in the longer retraining comparison (Methods).

Figures~\ref{fig:3}a and~\ref{fig:3}b show the test-set RMSE on energy and on force respectively for the four picker strategies across the K-sweep. At the smallest label count K = 50, the three uncertainty-driven pickers already largely outperform random sampling on the bulk metrics. On mean E (Figure~\ref{fig:3}a), the model committee is the strongest of the four pickers at about 2.2 meV/atom, followed by energy LLPR at about 3.1 and force LLPR at about 6.6, while random sampling lags behind at about 11.9 meV/atom. On mean F (Figure~\ref{fig:3}b) the same ordering holds, with committee at about 0.05 eV/\AA{}, energy LLPR at about 0.06 eV/\AA{}, force LLPR at about 0.07 eV/\AA{}, and random at about 0.12 eV/\AA{}. At K = 100, all three uncertainty-driven pickers continue to outperform random sampling and the LLPR pickers catch up with committee: energy LLPR reaches a mean E of about 1.7 meV/atom against committee's 2.3, both approaching the full-data ceiling of 1.30 meV/atom; on mean F all three uncertainty-driven pickers are within 9\% of the 0.043 eV/\AA{} ceiling, while random sits at about 0.06 eV/\AA{}. From K = 150 to K = 200 the uncertainty-driven pickers plateau close to the ceiling, while random keeps improving with K but does not fully close the gap.

This contrast indicates that the uncertainty-driven strategies efficiently cover the structurally informative regions of the candidate pool by about K = 100, whereas random sampling needs more frames to achieve comparable coverage. At medium-to-large K the LLPR pickers are also slightly ahead of the model committee on the mean metrics, indicating that the LLPR uncertainty signal carries comparable or stronger discriminative information than the committee spread for picking, and at a substantially lower training cost, since building a model committee requires training multiple independent models whereas LLPR only requires one. Notably, at K = 100 the LLPR-driven workflow already reaches a mean test E of about 1.7 meV/atom and a mean test F of about 0.047 eV/\AA{} using 300 configurations for retraining (the 200-frame liquid anchor plus 100 picks), approaching the full-data ceiling on both energy and force; this confirms that near full-data accuracy can be reached with substantially fewer labels when those labels are selected by an informative UQ signal.

Figures~\ref{fig:3}c and~\ref{fig:3}d show the worst-1\% mean test E and worst-1\% mean test F respectively, i.e.\ the mean of the largest nine per-frame errors out of 990. This metric is more discriminating than the bulk mean E and mean F because it isolates exactly the rare hard frames where the model breaks down. The contrasts between strategies are accordingly much sharper. At K = 50, random reaches a worst-1\% mean E of about 103.7 meV/atom and a worst-1\% mean F of about 0.80 eV/\AA{}, whereas the model committee reaches about 9.8 meV/atom and 0.095 eV/\AA{}, an order of magnitude lower on energy and a factor of eight lower on force. The two LLPR pickers are intermediate at K = 50 and reach committee-level worst-1\% performance by K = 100. All three uncertainty-driven pickers converge to within a factor of two of the full-data ceiling (5.4 meV/atom on worst-1\% mean E, 0.102 eV/\AA{} on worst-1\% mean F) by K = 100, while random sampling is still about 7 times above the ceiling on energy and 3 times above on force at the same K.

A further practical benefit of the uncertainty-driven pickers is reproducibility: their observed cross-seed standard deviations on the worst-1\% mean are smaller than random sampling's in this comparison, consistent with uncertainty ranking concentrating the picks on informative frames (for example, at K = 100 the cross-seed standard deviation on worst-1\% mean E is about 1--2 for the LLPR pickers and the committee, against about 40 for random). LLPR matches the committee's reproducibility across seeds while using a single model for uncertainty estimation.

\subsection*{Extending the workflow to foundation MLFF fine-tuning}

The from-scratch case above demonstrated the LLPR-based data-selection effect on a model trained from random initialisation. In practice, deploying an MLFF to a new system is increasingly done by fine-tuning a foundation MLFF\cite{ref9,ref10,ref11,ref43}. We next investigate whether the same data-selection effect extends to the fine-tuning regime. We use the same Morawietz water/ice dataset, expand the candidate pool to 2140 frames (1140 ice and 1000 liquid frames), and replace the from-scratch DeePMD bootstrap with the MACE-MP-0 foundation\cite{ref10}. MACE-MP-0 small is used as the primary foundation, and MACE-MP-0 medium is included for a model-size comparison. The fine-tuning recipe uses the multihead approach; full hyperparameters (replay buffer, loss weights, learning rate, etc.) are in Methods. As in the from-scratch case, K denotes the cumulative number of DFT-labelled frames acquired by the active-learning workflow. We sweep K from 50 up to 200 in steps of 50 with three retraining seeds per cell. A model-committee baseline is not presented in this regime: constructing one would require multiple independent foundation fine-tuning runs, which is computationally impractical.

\begin{figure}[!htbp]
\centering
\includegraphics[width=\textwidth]{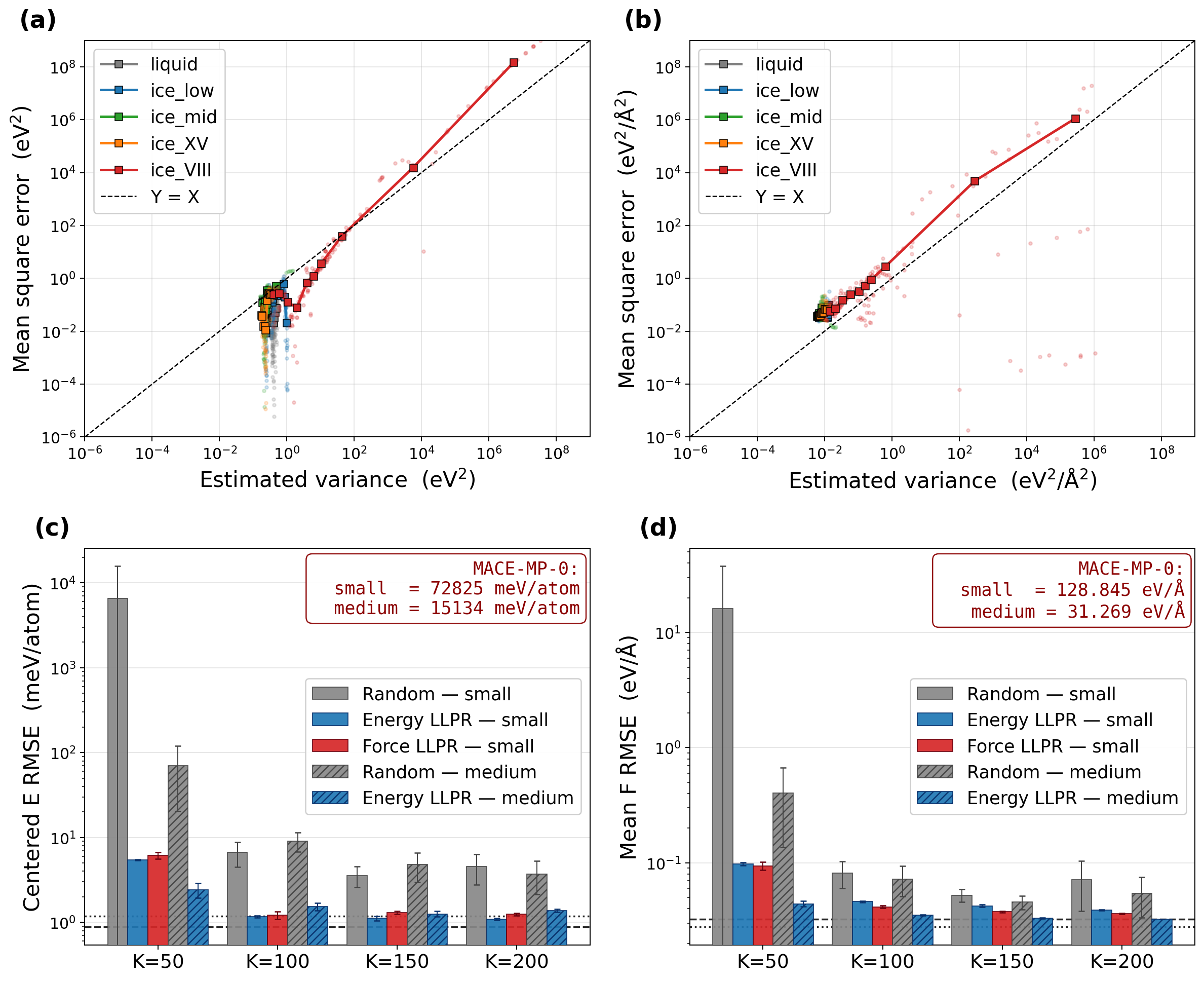}
\caption{\textbf{LLPR-based picking approaches the foundation fine-tuning reference on water/ice with far fewer labels.} \textbf{(a, b)} LLPR-based uncertainty of MACE-MP-0 small on the candidate pool: predicted energy uncertainty $\sigma^2_E$ against the constant-offset-corrected ground-truth squared energy error (a); predicted top-5 force uncertainty $\sigma^2_F$ against the corresponding top-5 squared force error (b). Each subsystem (liquid, ice\_low, ice\_mid, ice\_XV, ice\_VIII) is plotted in a distinct colour with per-phase bin-mean curves; the dashed line is the y = x diagonal. \textbf{(c, d)} K-sweep of three picker strategies (random, energy LLPR, force LLPR top-5) on the MACE-MP-0 small foundation (solid bars), and two strategies (random, energy LLPR) on the MACE-MP-0 medium foundation (hatched bars), at K = 50, 100, 150, 200 cumulative DFT-labelled picks with three independent retraining seeds per cell: subsystem-centered test energy RMSE (c) and unshifted force RMSE (d). The dashed and dotted horizontal lines in each panel are the full-data fine-tuning references for the small and medium foundations, respectively, obtained by fine-tuning each foundation on the 3955-frame reference training set under the same recipe. The bare MACE-MP-0 mean test errors for both foundations are given as the annotation in the upper right of each panel. Error bars are population standard deviations across the three retraining seeds (divisor $n=3$).}
\label{fig:4}
\end{figure}

Figure~\ref{fig:4}a--b show the LLPR-based uncertainty estimation on MACE-MP-0 small. The energy and top-5 force uncertainties are evaluated against their ground-truth squared errors on the candidate pool combined with an in-distribution liquid probe drawn from the bootstrap training set (details in Methods). The bin-mean energy and force curves compare the uncertainty scale with observed squared error across this pool. Energy errors use the signed predictions and a composition offset fitted on liquid training anchors; the calibration plot is descriptive and does not establish exact agreement or coverage on an independent test distribution. Notably, almost all of the high-uncertainty frames come from a single subsystem, ice\_VIII, the high-pressure cubic ice polymorph with density above 1.5 g/cm$^3$, while the in-distribution liquid frames sit at the low-variance end of the plot with the smallest uncertainties of any phase, indicating that the foundation is already near-quantitative on the in-distribution part of the target chemistry (see SI Figure~S1 for per-subsystem zero-shot RMSEs).

Figure~\ref{fig:4}c--d turn the LLPR-based uncertainty into a data-selection rule and report subsystem-centered test energy RMSE and unshifted force RMSE. The energy metric differs from the absolute energy RMSE used in the from-scratch case (see Methods). Two reference points anchor each panel. The bare MACE-MP-0 foundations, evaluated zero-shot without any fine-tuning, give a mean test energy of 72{,}825 meV/atom on small and 15{,}134 meV/atom on medium, and a mean test force of 128.8 eV/\AA{} on small and 31.3 eV/\AA{} on medium (annotation, upper right of each panel). The medium foundation is several-fold better than small in absolute terms before fine-tuning, reflecting its larger architectural capacity. The per-subsystem breakdown (Figure S1) shows that these aggregate numbers are dominated almost entirely by ice\_VIII; the median subsystem error is below 5 meV/atom on both foundations, consistent with the uncertainty distribution in Figure~\ref{fig:4}a--b. The dashed (small) and dotted (medium) horizontal lines in each panel mark the full-data fine-tuning references: 0.89 meV/atom and 1.18 meV/atom on mean test energy for small and medium respectively, and 0.032 eV/\AA{} and 0.028 eV/\AA{} on mean test force, obtained by fine-tuning each foundation on the same 3955-frame Morawietz reference training set under the same recipe.

With K = 50 random picks, the small foundation retains a large mean test energy error of about 6550 meV/atom (Figure~\ref{fig:4}c). The observed random-selection performance at this budget varies strongly across seeds. With the same label budget, the medium foundation reaches a mean test energy error of about 70 meV/atom, roughly two orders of magnitude lower. By K = 100 the random gap between small and medium has effectively closed (mean test energy of 6.66 and 9.12 meV/atom respectively); however, both numbers are within roughly one order of magnitude of the corresponding full-data fine-tuning references. The LLPR-based pickers behave very differently. At K = 50 the energy-LLPR picker on the small foundation reaches a mean test energy of 5.4 meV/atom, already three orders of magnitude below the random baseline at the same K, and at K = 100 it reaches 1.17 meV/atom, close to the small foundation's full-data fine-tuning reference of 0.89 meV/atom. The force-LLPR picker behaves comparably at the same K (1.21 meV/atom on mean test energy at K = 100). For K $\ge$ 100 the errors of the LLPR-based pickers plateau near their respective full-data fine-tuning references; doubling the label budget from K = 100 to K = 200 reduces the mean test energy from 1.17 to 1.09 meV/atom on the small foundation with energy LLPR, a marginal further improvement. The same pattern carries through to force errors. The LLPR-based pickers reach a mean test force of approximately 0.04 eV/\AA{} at K = 100 on the small foundation (close to its full-data fine-tuning reference of 0.032 eV/\AA{}), while random picking plateaus near 0.08 eV/\AA{} at K = 100 and 0.07 eV/\AA{} at K = 200, remaining roughly a factor of two above the full-data fine-tuning reference.

These findings indicate that, in this water/ice test, \emph{which frames} are picked matters more than \emph{which foundation size} is used. The K = 50 energy-LLPR picks on the small foundation (mean test energy about 5.4 meV/atom) already beat the K = 100 random picks on the medium foundation (mean test energy 9.12 meV/atom), despite the small foundation being substantially weaker before fine-tuning, and with K = 100 LLPR-based picks both foundations are close to their respective full-data fine-tuning references on both energy and force (Figure~\ref{fig:4}c--d).

The cross-seed error bars at low K are also dramatically narrower under LLPR-based picking than under random picking, paralleling the from-scratch case. For example, at K = 50 on the small foundation, the energy-LLPR run gives a cross-seed standard deviation of 0.06 meV/atom on mean test energy against random's 9196 meV/atom, about five orders of magnitude smaller. The data-selection effect demonstrated in the from-scratch case therefore reproduces in the fine-tuning case. Roughly 100 LLPR-picked DFT-labelled frames bring both the small and medium foundations close to their full-data fine-tuning references on energy and force, whereas random picking at the same K retains larger errors.

\subsection*{Detecting and resolving foundation MD hallucination in active fine-tuning}

Previously we used the open Morawietz water/ice dataset\cite{ref26,ref27} as a controlled benchmark, in which the DFT-labelled candidate pool is available in advance. In practice this is rarely the case. For a new target system, no labelled data are available and the fine-tuning data must be constructed on the fly. Here we implement the LLPR-based active-learning workflow on the 21 molal LiTFSI aqueous electrolyte, a popular electrolyte material that is structurally challenging because the salt mass exceeds the solvent mass at this concentration, hence the name \emph{water-in-salt} electrolyte\cite{ref22}. In this test we do not assume any prior labelled samples for the target system.

We deliberately choose MACE-OMAT small\cite{ref9,ref49} as the foundation. The model is pretrained on the OMAT24 open materials database and has minimal exposure to dense aqueous configurations. It is also one of the smallest MACE foundation models. Two reasons motivate this choice. First, a crystalline-trained foundation deployed on a dense aqueous electrolyte is a foundation-target chemistry mismatch that is common in realistic MLFF deployment, and we want to see how far fine-tuning can close the gap that pretraining left. Using a small foundation allows us to focus on the effect of data selection.

\begin{figure}[!htbp]
\centering
\includegraphics[width=\textwidth]{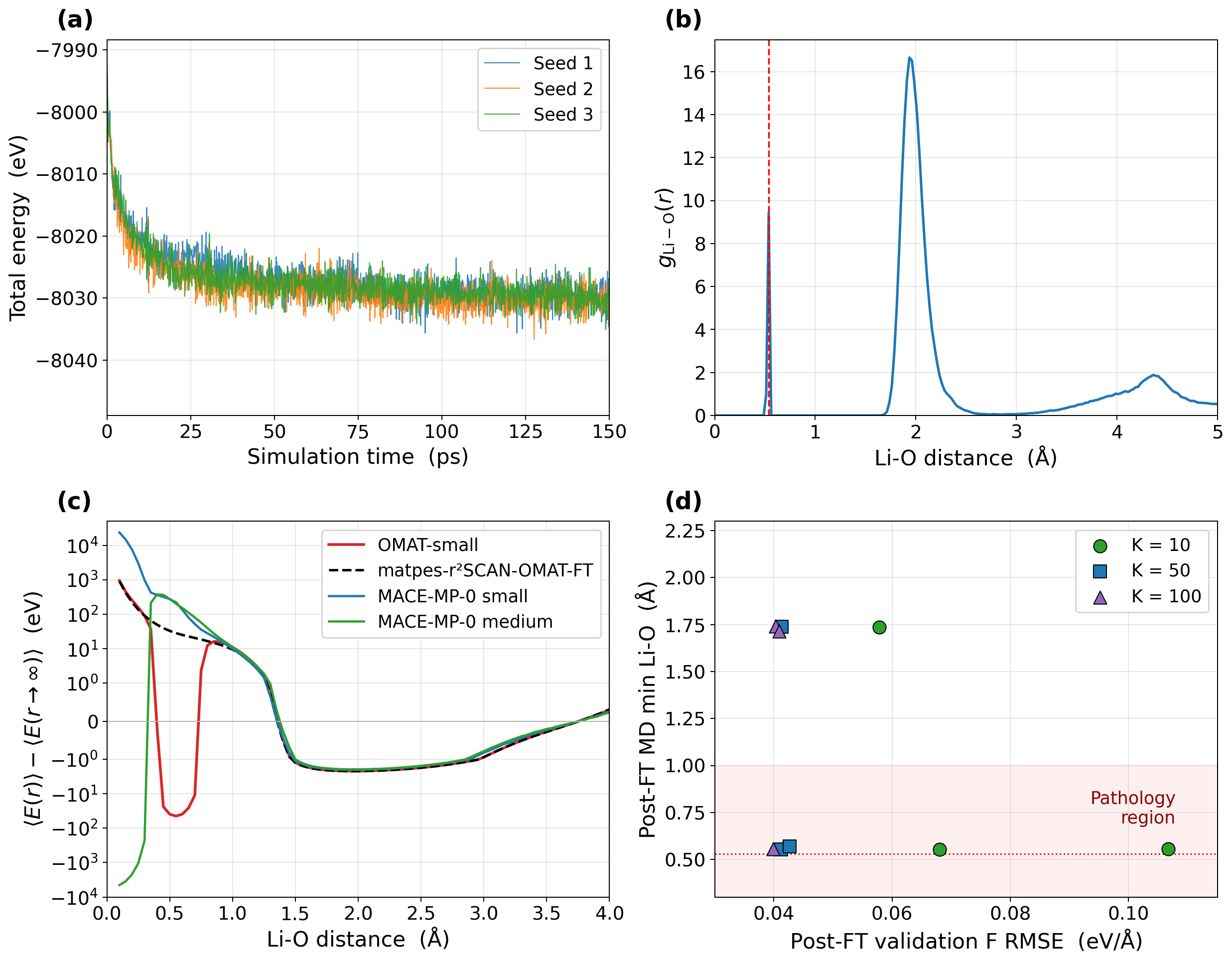}
\caption{\textbf{Foundation MD hallucination on 21 m LiTFSI electrolyte: stable but structurally wrong, and standard fine-tuning diagnostics do not detect it.} \textbf{(a)} Total energy over the first 150 ps of three bare OMAT-small NPT production records at 300 K. \textbf{(b)} Bare-model Li--O radial distribution function. The inset is a real-snapshot rendering of the local cluster around the trapped \ce{Li+} ion. \textbf{(c)} Multi-frame rigid-geometry Li--O potential-energy scan for four different foundation models. \textbf{(d)} Final-20-ps mean of the instantaneous minimum Li--O distance versus final validation force RMSE (eV/\AA{}) for the nine one-shot fine-tunings (K = 10, 50, 100 random picks with 3 retraining seeds each).}
\label{fig:5}
\end{figure}

We first run three independent 100 ps NPT simulations at 300 K with the bare OMAT-small foundation on the 21 m LiTFSI electrolyte configuration. The trajectories are stable by standard MD diagnostics. The plotted total energy remains bounded over the first 150 ps of the longer archived bare-model production records (Figure~\ref{fig:5}a), and no numerical instability occurs. The density converges to 1.59 g/cm$^3$, about 7\% below the experimental value of 1.7126 g/cm$^3$ for 21 m LiTFSI at 298 K\cite{ref46}. The radial distribution function of Li--O, however, shows a persistent sub-angstrom peak that should not be there (Figure~\ref{fig:5}b): a single \ce{Li+} ion is locked at a 0.53 \AA{} contact with an adjacent water O (see Figure~\ref{fig:5}b inset) throughout the trajectory, while the physically expected first-shell peak at 1.97 \AA{} is unchanged. The Li--O potential-energy scan (see Methods) in Figure~\ref{fig:5}c traces the cause: the OMAT-small PES has a hallucinated attractive well at 0.55 \AA{}, roughly 45 eV below its long-range value. This is probably because Li--O contacts below $\sim$1.5 \AA{} are essentially absent from the OMAT24 pretraining set\cite{ref9}, so the network freely extrapolates its learned Li--O attractive prior into the unsampled short-range region without any architectural constraint forcing repulsion.

Previous benchmarking of foundation MLFFs on 21 m LiTFSI has shown that the matpes-r$^2$SCAN-OMAT-FT foundation accurately reproduces the experimental thermodynamic and dynamical properties of this electrolyte\cite{ref47}, and its scanned Li--O PES lacks the sub-angstrom attractive well (Figure~\ref{fig:5}c). We therefore adopt it as a pseudo-oracle to label active-learning samples instead of running DFT. The goal here is not to develop a production force field for 21 m LiTFSI but to test whether a hallucinating foundation can be repaired by fine-tuning. To test whether standard fine-tuning diagnostics can detect the OMAT-small hallucination, we perform one-shot fine-tunings on K = 10, 50, and 100 random picks from the foundation's hallucinated MD trajectory, labelled with matpes-r$^2$SCAN-OMAT-FT, with three retraining seeds per K. The fine-tuning recipe follows the same multihead template as before (details in Methods). Figure~\ref{fig:5}d plots the nine fine-tuned models by final validation force RMSE (the standard fine-tuning diagnostic, $x$-axis in eV/\AA{}) against the mean of the instantaneous minimum Li--O distances over the final 20 ps of each post-fine-tuning MD record (a deployment-quality probe). The K = 10 fine-tunes span a wider range of validation force RMSE (0.058--0.107 eV/\AA{}), as expected at the data-limited end of the K-sweep. By K = 50 and K = 100, the six fine-tunes improve to a validation force RMSE of approximately 0.040 eV/\AA{}, which would conventionally be considered accurate enough. The post-fine-tuning MD outcomes, however, are inconsistent even at K = 100: in five of the nine retraining runs the model inherits the foundation pathology despite low validation errors. The outcome varies across seeds even at a fixed label budget. These nine tests show that a low validation force RMSE alone does not guarantee removal of the short-range Li--O feature.

\begin{figure}[!htbp]
\centering
\includegraphics[width=\textwidth]{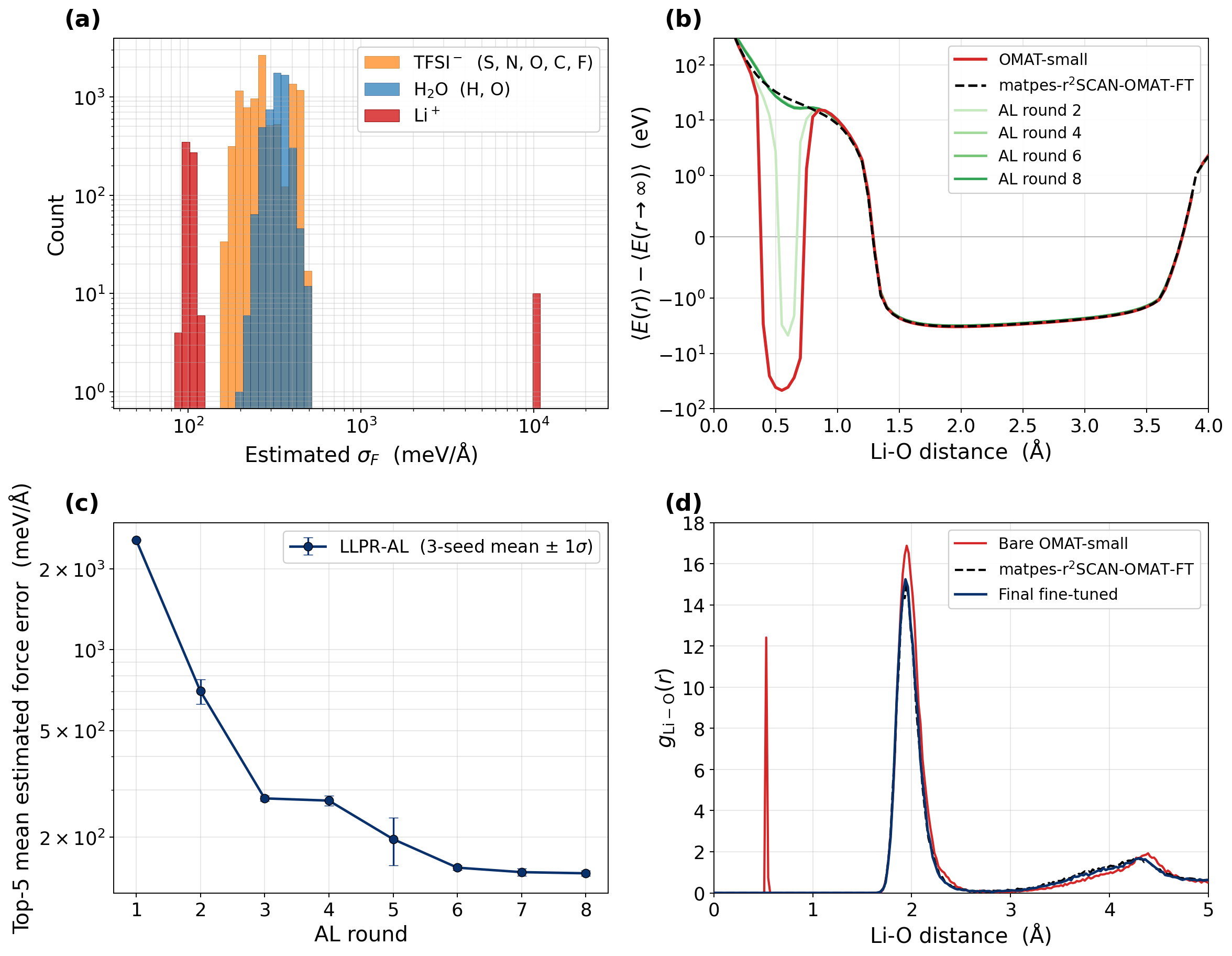}
\caption{\textbf{LLPR-based active fine-tuning mitigates the short-range Li--O feature in 21 m LiTFSI.} \textbf{(a)} Atomic force-uncertainty scale $\sigma_F$ in meV/\AA{}, comprising 15,340 values from 10 frames of one bare-model trajectory, separated by molecular species. \textbf{(b)} Mean rigid Li--O scans on five shared backgrounds for seed 42; each curve is re-zeroed using $r>3.5$ \AA{}. \textbf{(c)} Mean over frames of the five largest atomic $\sigma_F$ values, averaged over three seeds with population-standard-deviation error bars. Round $k$ evaluates $M_{k-1}$ using calibration on the next ten selected frames. \textbf{(d)} Li--O RDF after eight active-learning rounds (three-seed mean, final 50 saved frames per trajectory), with bare-model and pseudo-oracle reference curves.}
\label{fig:6}
\end{figure}

The LLPR-based uncertainty supplies the missing detection signal that the validation loss cannot. Computing the per-atom force uncertainty of the bare OMAT-small foundation on its hallucinated-MD trajectory, we find that the bulk of the atomic uncertainty distribution is well below the Li outlier tail (median $\sigma_F \approx 300$ meV/\AA{}); the exceptions are the Li atoms locked at the pathological 0.53 \AA{} contact, whose $\sigma_F$ extends in a long tail to roughly 10{,}000 meV/\AA{} (Figure~\ref{fig:6}a). The signal distinguishes a configuration on which the foundation is interpolating from one on which it is extrapolating, which the validation-set loss cannot do because the validation set is drawn from the same hallucinated MD as the training set. The leverage ranking requires no new target-system labels. The histogram's absolute scale additionally uses an arithmetic calibration factor from the pretraining-domain reference data; it is distinct from the per-round geometric calibration below.

We then run iterative LLPR-based active learning with the following steps. Each round picks 10 frames from the current MD ranked by the mean of the five largest per-atom force uncertainties, labels them with the matpes-r$^2$SCAN-OMAT-FT pseudo-oracle, accumulates them with the previous picks, and fine-tunes from the OMAT-small foundation with the multihead recipe (see Methods). We assess models through round 8 for three retraining seeds. The workflow uses a prescribed maximum round count, with calibration performed separately. Figure~\ref{fig:6}b shows the seed-42 Li--O potential-energy scan (the other two seeds are shown in Supplementary Figure S2), comparing the bare OMAT-small foundation against the LLPR-based active-learning fine-tunes at rounds 2, 4, 6 and 8 alongside the matpes-r$^2$SCAN-OMAT-FT reference. By round 2 the hallucinated well has been substantially lifted but a shallow residual attraction at $\sim$0.7 \AA{} remains; by round 4 the sub-angstrom attractive minimum has disappeared and a steep short-range repulsive wall has emerged. The curves at rounds 6 and 8 are essentially indistinguishable.

Figure~\ref{fig:6}c plots the per-round per-frame top-5 mean uncertainty averaged across the three retraining seeds. It falls rapidly from approximately 2564 meV/\AA{} at round 1 to about 700 meV/\AA{} at round 2, 280 meV/\AA{} at round 3, and stabilises at approximately 150 meV/\AA{} from round 6 onwards. Each point indexed $k$ evaluates the generating model $M_{k-1}$, before training on the round-$k$ picks. The decline supports retrospective convergence monitoring. Calibration was performed retrospectively; the simulations used a fixed maximum number of rounds. The separate atomic 99th percentile first falls below the proposed 100 meV/\AA{} threshold at round 4 for all three seeds, while the top-5 mean remains larger. Figure~\ref{fig:6}d shows the Li--O radial distribution function from production NPT trajectories at 300 K and 1 atm comparing the bare OMAT-small foundation, the fine-tuned model at round 8 (3-seed mean), and the matpes-r$^2$SCAN-OMAT-FT reference. The pathological 0.53 \AA{} peak is gone in the fine-tuned model, and its Li--O RDF closely tracks the matpes-r$^2$SCAN-OMAT-FT reference across the full range. The three-seed mean density over the final 50 ps of the trajectories generated by the models after eight active-learning rounds, $1.705 \pm 0.044$ g/cm$^3$, agrees with both the matpes reference value of $1.720$ g/cm$^3$ and the experimental value of $1.7126$ g/cm$^3$\cite{ref47}, showing agreement in these selected structural and thermodynamic observables. Taken together, the workflow mitigates the targeted short-range Li--O pathology in the 21 m LiTFSI system.

\section*{Discussion}
Data efficiency is central to machine-learning force field training. In our water/ice benchmarks, LLPR selection approaches full-data reference accuracy with fewer additional target-system labels in both train-from-scratch and foundation fine-tuning workflows. These label budgets refer to newly selected target configurations, in addition to the bootstrap or pretraining and replay data. Although a pretrained foundation already offers good out-of-the-box accuracy on many systems, deploying it on a new target chemistry still demands a carefully constructed fine-tuning dataset to cover the configurations the target system samples in MD. Beyond the label-cost concern, pathological extrapolation is particularly relevant to the foundation paradigm. It arises when the target chemistry falls outside what the foundation has seen during pretraining, and these failures can persist silently into deployment if not specifically probed. A model committee is commonly used in active learning to flag such pathologies, but it requires training $N$ independent MLFFs and is therefore costly even in the train-from-scratch setting; foundation deployment adds the cost of an independent fine-tuning run for each committee member. We show in this work that the last-layer-projection-regression (LLPR) framework\cite{ref20,ref45}, equipped with a target-relevant covariance and round-wise calibration in force-error units, provides a lightweight forward-pass uncertainty signal that addresses both concerns across the three deployment regimes we examined and achieves full-data accuracy with substantially fewer DFT labels in both the train-from-scratch and the foundation fine-tuning settings.

In the train-from-scratch setting on a structurally heterogeneous water/ice candidate pool, the LLPR-driven workflow reaches the full-data-trained ceiling using only K = 100 labelled picks, an order of magnitude fewer labels than the 3955-frame full-data baseline. The LLPR picker is roughly twice as sample-efficient as random and competitive with the model committee at substantially lower compute, with cross-seed variance tight enough that picks made with a single trained model match the reproducibility of those from a multi-seed committee.

In the foundation fine-tuning setting on the same water/ice dataset, the picker effect transfers cleanly. Energy-LLPR picking on the MACE-MP-0 small foundation approaches its full-data fine-tuning reference at K = 100, while random picking retains larger errors at the same K. Crucially, the bare performance gap between MACE-MP-0 small and MACE-MP-0 medium (more than fourfold at zero-shot) closes within K = 100 LLPR picks, showing that, in this test, which frames are labelled matters more than which foundation is used.

In the realistic deployment regime where no pre-labelled target-system pool is available (the 21 m LiTFSI electrolyte case), LLPR plays a different role. The bare OMAT-small foundation produces molecular dynamics that is stable by standard MD diagnostics yet structurally wrong, with a \ce{Li+} ion locked at a 0.53 \AA{} contact with water that the foundation's own PES has hallucinated as an attractive well. The LLPR per-atom uncertainty flags this hallucination directly on the affected frames before any labels are committed, drives the iterative active-learning loop, and supports retrospective monitoring after calibration against pseudo-oracle labels. The fine-tuned models improve Li--O coordination and reproduce the reported density.

By integrating last-layer-projection regression uncertainty into the active-learning workflows used to train and fine-tune MLFFs, we show that full-data accuracy can be reached with substantially fewer labels and that the model's own internal signal can drive an end-to-end workflow. The workflow is implemented in the open AI$^2$-kit\cite{ref30} active-learning framework and supports both train-from-scratch and foundation fine-tuning pipelines; it is currently demonstrated on the MACE foundation family but the per-atom-feature hooks on which it relies generalise straightforwardly to other equivariant foundation architectures. Compared with the model committee, which requires $N$ independently trained MLFFs and returns an uncertainty that is only relative across members, LLPR provides a quantitatively accurate, deterministic uncertainty from a single trained model, matching the committee on picker performance at a fraction of the compute. LLPR thus provides a single-model uncertainty signal for reducing target-system labelling requirements, identifying unphysical configurations during MD, and assessing convergence retrospectively.

\section*{Methods}

\subsection*{LLPR validation on bulk water}

The bulk-water experiment uses four DeePMD models and a 50-configuration reference set sampled from the dataset of ref.~\cite{ref48}. The training recipe assigns 45 configurations to training and five to validation, using the smooth-edition descriptor, a [240, 240, 240] fitting network and 10000 optimisation steps. The exponential learning-rate schedule starts at $10^{-3}$ with a limiting rate of $10^{-7}$.

Evaluation uses 1646 unique configurations excluding the 50-configuration reference set: 585 with 64 waters, 1022 with 63 waters and 39 with 128 waters. All methods use this same evaluation pool. The full reference set is used for LLPR covariance and uncertainty-scale calibration; the random-retrieval baseline averages 200 permutations.

For each configuration $x$, the last-layer activations of the DeePMD fitting network are hooked to produce a feature vector $h_a(x)$ per atom $a$ (sum-pooled per frame for the energy signal). Given a reference set $\{x_i\}$, the per-atom feature covariance $\Sigma_a$ is accumulated and regularised to give a regularised covariance matrix $A$; the per-atom leverage $u_a(x)$ then measures how far the atomic feature lies outside the training distribution:
\begin{equation}
\Sigma_a = \sum_{i,a} h_a(x_i)\,h_a(x_i)^\top, \quad
A = \tfrac{1}{N_\text{tot}}\Sigma_a + \sigma_F^2 I, \quad
u_a(x) = h_a(x)^\top A^{-1} h_a(x),
\label{eq:llpr}
\end{equation}
with the Tikhonov ridge $\sigma_F = 5 \times 10^{-3}$ throughout. The energy-LLPR uncertainty $\sigma^2_E(x)$ is defined analogously on the sum-pooled per-frame features against a per-frame covariance, and the force-LLPR uncertainty is the per-frame mean of the five largest per-atom $u_a$.

\subsection*{Train-from-scratch active learning on water/ice}

The water/ice dataset\cite{ref26,ref27} (7241 RPBE-D3 configurations) is partitioned as follows: the bootstrap MLFF$_0$ is trained on 1000 liquid-water frames; the active-learning candidate pool is the 1140-frame ice partition; the multi-phase test set is 990 frames (500 liquid + 490 ice); the from-scratch full-data ceiling trains on a 3955-frame union of the bootstrap, the active-learning pool, and the remaining small-cell liquid and ice frames not held out for testing. The 2296 large-cell frames are reserved as additional cell-size OOD probes (Table S2). The DeePMD bootstrap and retraining configurations for this separate water/ice campaign are provided with its reproduction scripts.

Four picker strategies are compared: random sampling, energy LLPR (top-K by $\sigma^2_E$), top-5 force LLPR, and model committee. The acquisition driver selects five new frames per round, producing cumulative budgets $K \in \{50, 100, 150, 200\}$. Its initial committee uses four bootstrap models, and later scoring uses the three retraining-seed models. The final budget comparison uses the 500000-step retraining recipe. Each retrained model is warm-started from the bootstrap and trained on a 200-frame liquid anchor plus the top-K active-learning picks. Three independent retraining seeds are run per (picker, K) cell.

\subsection*{Foundation fine-tuning on water/ice}

The bootstrap MLFF$_0$ is replaced by MACE-MP-0 small or medium\cite{ref10}, retaining the same 990-frame test set and 3955-frame full-data reference. The selection pool contains 2140 configurations (1140 ice and 1000 liquid frames). Uncertainty-guided selection combines LLPR scores with SOAP-based farthest-point sampling\cite{ref24}. Multihead fine-tuning uses target labels together with a 7000-frame H/O-only MPtrj replay partition; a separate 500-frame partition is used for covariance calibration. Energy and force loss weights are both 100. AdamW optimisation uses a learning rate of $10^{-4}$, EMA decay 0.99 and weight decay $5\times10^{-7}$ for 20 epochs per budget.

For the calibration analysis, a composition-proportional energy offset is fitted using ten random ten-frame subsets of a fixed 200-frame liquid training anchor and averaged across subsets. For the K-sweep and zero-shot energy metrics, each test subsystem's mean energy residual is subtracted before computing its RMSE. These centered errors measure within-subsystem variation rather than relative mean energies between phases; forces are unshifted. Raw energy errors and errors aligned using a training-only anchor are reported in Table S3.

\subsection*{Iterative LLPR-based active learning on 21 m LiTFSI electrolyte}

The target system is a 21 molal LiTFSI water-in-salt configuration: 1534 atoms per cell (64 \ce{Li+}, 64 TFSI$^-$, 170 H$_2$O). Labels are produced by the matpes-r$^2$SCAN-OMAT-FT foundation, which has been independently benchmarked on 21 m LiTFSI and reproduces the experimental density, ion/solvent diffusion coefficients, and X-ray structure factor\cite{ref47}.

\emph{Calibration.} The ten selected frames of round $k$ have not yet been used to train the generating model $M_{k-1}$, but are selected by its uncertainty and therefore are not an unbiased test set. On these picks we fit the geometric-mean estimator
\begin{equation}
C_\text{geom} = \exp\!\left(\langle \log(e_a^2 / u_a) \rangle_a\right), \qquad
\sigma_F(a;x) = \sqrt{C_\text{geom}\cdot u_a(x)},
\label{eq:Ccalib}
\end{equation}
where $e_a$ is the per-atom force error of the previous round's model against the pseudo-oracle labels, $u_a$ is the LLPR leverage of Eq.~\ref{eq:llpr}, and $\sigma_F$ is the resulting absolute per-atom force-error uncertainty in eV/\AA{}. Calibration is performed after sampling. The atomic 99th percentile is compared retrospectively with a proposed threshold of 100 meV/\AA{}.

\emph{Molecular dynamics and Li--O scans.} Three independent bare-model NPT tests use 100 ps trajectories with a 0.5 fs timestep at 300 K and 1 atm. The Nos\'{e}--Hoover thermostat relaxation time is 0.05 ps; the Melchionna barostat\cite{ref50} relaxation time is 0.75 ps with a reference bulk modulus of 2.2 GPa. Figure~\ref{fig:5}a shows the first 150 ps of separate, longer production records. Rigid Li--O scans displace one Li along the Li--O axis with all other atoms fixed, evaluating 79 distances from 0.10 to 4.00 \AA{} and averaging five backgrounds. Figure~\ref{fig:6}b uses five evenly spaced round-10 backgrounds, shared across compared models within each seed. Curves are shifted by the mean energy above 3.5 \AA{}.

\emph{One-shot and iterative active learning.} OMAT-small fine-tuning uses a 20,000-frame Li-bearing OMAT24 replay buffer. The nine one-shot runs use $K\in\{10,50,100\}$ target labels, three seeds, 30 epochs, batch size 4 and a 10\% target validation fraction. The covariance reference uses 2000 Li-bearing frames in round 1 and accumulated active-learning configurations thereafter. Each iterative round acquires ten labels and fine-tunes from the foundation for five epochs. Three seeds are run through ten rounds. After eight rounds, each model has acquired 80 target configurations. Algorithm~\ref{alg:executed} summarizes the workflow. Calibration is evaluated separately from the sampling and training loop.

\begin{algorithm}[htbp]
\caption{Electrolyte active-learning workflow and retrospective calibration.}
\label{alg:executed}
\begin{algorithmic}[1]
\Require Foundation $M_0$, pseudo-oracle, replay set, initial coordinates, maximum round count
\State Set acquired data $D=\varnothing$ and current model $M=M_0$.
\For{each round $k$ through the prescribed maximum}
\State Run 10 ps NPT MD; retain 100 snapshots across the full trajectory at 100 fs spacing.
\State Build the covariance from 2000 Li-bearing reference frames if $k=1$, otherwise from $D$.
\State Rank frames by the mean of their five largest atomic leverages.
\State Greedily select ten frames from the top 20\%, applying a cosine-similarity cutoff of 0.98 to normalized sum-pooled features.
\State If fewer than ten survive, fill the batch by rank from the remaining top-20\% candidates.
\State Label with the pseudo-oracle and append to $D$.
\State Fine-tune from $M_0$ on $D$ with pretraining replay (5 epochs, learning rate $10^{-4}$, EMA decay 0.99); set $M$ to the result.
\State Save the model, picks, final MD coordinates and scan diagnostic.
\EndFor
\State Separately, calibrate each generating model $M_{k-1}$ on its round-$k$ picks and summarize uncertainty on its sampled frames.
\end{algorithmic}
\end{algorithm}

\section*{Acknowledgements}

This work was supported by the New Generation Artificial Intelligence---National Science and Technology Major Project (2025ZD0122703), the National Natural Science Foundation of China (Grant Nos. 22503072 and 22225302), and the Fundamental Research Funds for the Central Universities (20720250122).

\section*{Author contributions}

S.B.: Conceptualization, Software, Investigation, Writing---original draft, Writing---review \& editing. Y.-Z.W.: Software, Investigation, Formal analysis, Writing---review \& editing. J.C.: Supervision, Writing---review \& editing.

\section*{Competing interests}

The authors declare no competing interests.

\section*{Data and code availability}

Supporting data, code, and analysis scripts are available to editors and reviewers upon request during peer review. After publication, supporting data will be made publicly available and the code and scripts will be released on GitHub.

\bibliography{references}

\end{document}